\documentclass{emulateapj} 
\usepackage{psfig}
\usepackage{aastexug}   

\begin{document} 
 
\title{The Color Selection of Quasars from Redshifts 5 to 10: Cloning and Discovery }
\author{Kuenley Chiu \altaffilmark{1,2}, Wei Zheng\altaffilmark{2}, Donald P. Schneider\altaffilmark{3}, Karl Glazebrook\altaffilmark{2},\\ Masanori Iye\altaffilmark{4}, Nobunari Kashikawa\altaffilmark{4}, Zlatan Tsvetanov\altaffilmark{2,5}, Michitoshi Yoshida\altaffilmark{6}, Jon Brinkmann \altaffilmark{7}  } 
 
\altaffiltext{1}{chiu@pha.jhu.edu}
\altaffiltext{2}{Department of Physics and Astronomy, Johns Hopkins University, 3400 North Charles Street, Baltimore, MD 21218}
\altaffiltext{3}{Department of Astronomy and Astrophysics, Penn State University, 525 Davey Lab, University Park, PA 16802}
\altaffiltext{4}{National Astronomical Observatory of Japan, 2-21-1 Osawa, Mitaka Tokyo, Japan 181-8588}
\altaffiltext{5}{National Aeronautics and Space Administration, Washington, DC 20546-0001}
\altaffiltext{6}{Okayama Astrophysical Observatory, National Astronomical Observatory of Japan, Honjo 3037-5, Kamogata, Asakuchi-gun, Okayama Japan. 719-0232}
\altaffiltext{7}{Apache Point Observatory, P.O. Box 59, Sunspot, NM 88349-0059}

\begin{abstract} 

We present simulations of quasar colors, magnitudes, and numbers at redshifts $5<z<10$ based on our discovery of ten new high-redshift quasars and the cloning of lower redshift Sloan Digital Sky Survey (SDSS) quasars.  The ten quasars have redshifts ranging from $z=4.7$ to $z=5.3$ and $i$-magnitudes of $20.21$ to $20.94$.  The natural diversity of spectral features in the cloned sample allows more realistic simulation of the quasar locus width than previously possible with synthetic template spectra.  Colors are generated for the $z>6$ epoch taking advantage of the new UKIDSS near-infrared filter set, and we examine the redshift intervals of maximum productivity, discussing color selection and survey depth issues.  On the basis of the SDSS sample, we find that the surface density of $z>4.7$ quasars increases by a factor of $3\times$ by extending 0.7 $i$-magnitudes deeper than the SDSS spectroscopic survey limit of $i=20.2$ -- correspondingly we predict a total of $\sim400$ faint quasars in the SDSS main area that have redshift $z>4.7$ and magnitudes $i<20.9$. 

\keywords{quasars: general; quasars: emission lines}

\end{abstract}

\section{Introduction}

In the past decade, the study of distant quasars has proliferated with the happy coincidence of large telescope availability, advanced imaging systems, and wide-area digital sky surveys such as the Sloan Digital Sky Survey (SDSS; York et al. 2000) and the Two Degree Field Quasar Survey (2QZ; Boyle et al. 2000).  These factors have grown our understanding of quasars from a few special objects into significant beacons of the distant universe.  The productive searches for quasars have fueled cosmological advances, among them, for example, the search for the epoch of hydrogen reionization in the distant universe.  With the SDSS discovery of luminous quasars at redshift $z\sim5$ (Fan et al. 1999, Zheng et al. 2000), and a leap to quasars at redshift $z\sim6$ soon after (Fan et al. 2001a), the uncertainty of this important epoch has been steadily reduced.   And through compilations of uniform catalogs of quasars, the distant and faint quasar luminosity function continues to be extended (SDSS, Schneider et al. 2003; 2QZ, Croom et al. 2004).  

Will new ground-based infrared surveys reproduce the spectacular quasar discoveries of the past five years?   How many quasars will deeper and redder surveys find?  As the next generation of near-IR surveys prepares to begin observations, these questions are of keen interest to survey teams and observers pursuing increasingly higher redshifts.  Cosmologists, too, speculate on how far back it will be possible to push the earliest formation of the dark halos necessary for quasar activity, and whether the first epoch of star formation and chemical evolution will someday be glimpsed.  

The schedule of new optical/infrared surveys suggests that $z>6$ objects will still be sparse for several more years.  The recently discovered SDSS $z\sim6$ objects represent only the brightest fraction of the population at this distance and are quite rare -- 12 discovered to date within $\sim 4500$ square degrees at $z$-band magnitudes $<20.2$ \citep{fan04}.   With no large-area fainter optical surveys planned for the near future, the best hope of increasing such quasar numbers appears to be through deep pencil-beam studies.  Deep imaging wells have been drilled in the optical, such as in the Hubble Ultra Deep Field, and near-infrared surveys like the UKIDSS (http://www.ukidss.org) also plan to dedicate time to such faint observations which may reach as deep as $J=25$.  

With few faint objects at high redshift to guide such observational efforts, valuable information about the high redshift quasar luminosity function as a function of magnitude may be elucidated from more abundant numbers at $z<6$.  The properties of lower redshift quasars can predict the limiting magnitude of the observations needed to find the first $z\sim7$ quasar and even beyond, filling in the statistical deficiencies accompanying such small populations.  Low redshift quasars can be used to effectively target yet unseen objects, a task which requires an understanding of expected color, magnitude, and completeness.  

The cloning of quasar spectra provides a tool to predict these parameters.  As the historical barriers of $z=4,5$, and 6 have been broken, these observational advances have confirmed a valuable result -- the spectra of quasars at high redshift look much like those of quasars nearby \citep{fan04}.   In this work, we exploit the similarity of near and distant quasars to simulate and predict the color properties of the next generation of objects at $z\sim7$ and above.  Past quasar selections have generally relied on composite quasar spectra.  Here, by contrast, we take advantage of the natural diversity of a low redshift population to preserve the possible range of high-redshift quasar colors.   We employ nearby quasars discovered in the SDSS to clone samples of higher redshift quasars, where the natural rarity of such objects makes analysis and prediction of properties otherwise difficult.  And with a sample of 10 newly discovered faint quasars at $z\sim 5$, we illustrate how such faint objects can provide magnitude and completeness information, unhindered by stellar contamination issues when coupled with semi-synthetic cloned spectra. 

\section{Simulating $z\sim 5,6$, and above: general method}

In this paper we discuss, in order of increasing redshift, three groups of quasars and their color simulations: 1) quasars at $z\sim 5$ which are discovered through $riz$ imaging of the SDSS, including 10 newly observed here, 2) quasars at $z\sim 6$ which are discovered through $iz$ imaging in the SDSS and independent $J$ imaging in the near infrared, and 3) quasars yet to be discovered at $z>6.5$ which will rely on a new infrared filter system such as the UKIDSS {\it ZYJHK}.  We focus on these three epochs because, as much by fortune as by design, the optical and infrared filter sets combined with the colors of stars and quasars conspire to make these redshifts of prime interest and productivity for quasar observers. 
 
\subsection{Quasar Color Behavior}

Simulations of quasar colors to date have generally relied on the redshifting of synthetic model spectra.  The color behavior of quasars with increasing redshift is simple to predict in a very general way.  The interstellar medium along the line of sight removes flux from any background source object, and as increasingly higher redshifts are probed, absorption from gas at each intervening redshift progressively damps the spectrum, becoming more severe with redshift.    Of relevance here is hydrogen Ly$\alpha$ at 1216\AA\,: as the Lyman-$\alpha$ break increases in wavelength with redshift and passes through neighboring photometric bands, the color difference ($m_{blue}-m_{red}$) can usually be expected to increase monotonically until the Lyman-$\alpha$ break reaches the effective end of the first filter.  

However, to calculate {\it accurate} color behavior requires modeling the quasar spectrum or the use of real quasar data.   And the prediction of quasar colors in three bands -- which is most useful for detailed target selection through color-color limits -- can be very sensitive to spectral features.  While diversity has been simulated in some synthetic models of quasars, including variations in emission and absorption lines as well as continuum properties, the natural diversity of quasars is much preferred.  

In each of these three redshift regimes, we produce semi-synthetic colors of distant quasars by using the known spectra of quasars at lower redshifts.  This procedure, called cloning, has been previously employed to simulate the images of high-redshift galaxies and aid photometric target selection, such as in Bouwens et al. (1998a,b).  Here we simulate the spectra of high redshift quasars by taking a large number of spectroscopically observed quasars at a lower redshift, numerically shifting those spectra to a desired higher redshift (where the real sample is otherwise sparse), and determining the resulting expected colors through selected broadband filters.  When a digital catalog of well-classified spectra is available, this technique is straightforward yet powerful and allows a simulation of quasar properties that need only be ``semi-synthetic''.  Diversity of spectral features is naturally introduced through the low redshift quasar sample itself, and propagated to higher redshift without large uncertainties or assumptions regarding the range of possible quasar attributes.  

We emphasize that the capability to clone quasars without having to consider intrinsic quasar evolution need not have been the case.  For example, galaxy spectra at high redshift may be significantly different from the galaxies which are familiar in the local neighborhood -- cloning such populations to high redshift actually yields an empirical measure of galaxy evolution when combined with observation.  Luckily, we are aided by the observation that quasars even at redshifts above $z=6$ display the same general spectral features as those at $z=2$, for instance.  

\subsection{Spectral Data Source and Cloning Procedure}

In this work, we use the SDSS DR1 Quasar Catalog generated by \citet{schneider} which covers 1360 square degrees and contains 16,713 quasars.  The DR1 Quasar Catalog is a compilation of quasars confirmed by the SDSS automated spectroscopic program, which reach $z=5.41$ in redshift and range from 15.15 to 21.78 in $i$-magnitude.  The source imaging data for this catalog was taken from MJD 51075 to 52522.  The hardware and software pipelines producing the final superb photometry and astrometry of the survey have been described by the project collaborators elsewhere in detail: \citet{york}, \citet{fukugita}, \citet{gunn98}, \citet{hogg}, \citet{lupton}, \citet{stoughton}, \citet{smith}, \citet{pier}, \citet{abz}.  

In each of the three redshift ranges, we began by selecting a subset of quasars from the DR1 Quasar Catalog.  These selections chose objects optimally suited for cloning -- the main criteria were that quasars would be of high enough redshift to display Lyman-$\alpha$, and that the cloned spectrum would have generally complete flux in the filters of interest.  It is worth recalling here that the raw SDSS spectra do not cover the full wavelength range of the SDSS imaging filter system -- {\it i.e.} the spectra are incomplete at the extremes of the $u$ and $z$ filters.  Thus during the generation of any synthetic photometry, care must be taken to ensure that spectral data exist down to the 3630\AA \, and up to 11230\AA \,(if data is desired in the $u$ and $z$ filters). 
 
Table 1 shows the number of quasars used in the each of the three cloning procedures, the filters covered, the redshift range from which quasars were extracted, and the redshift range of quasar clones produced.  

\begin{table}
\caption{Quasar cloning source data}
\begin{center}
\leavevmode
\begin{tabular}{ccccccc}
\hline \hline 
Cloned $z$ range & $N$ & Filters & Source $z$ range\\
\hline
$4.4-5.5$ & 741 & SDSS $riz$ & $2.5-3.3$ \\
$5.5-7.0$ &  756 & SDSS $iz$, UKIDSS $J$ & $2.83-4.4$ \\
$5.5-10.5$ & 1198 & UKIDSS {\it ZYJH} & $2.3-3.32$ \\
\end{tabular}
\end{center}
\end{table}

\begin{figure*}
\begin{center}
    \leavevmode
\epsfxsize=12cm\epsfbox{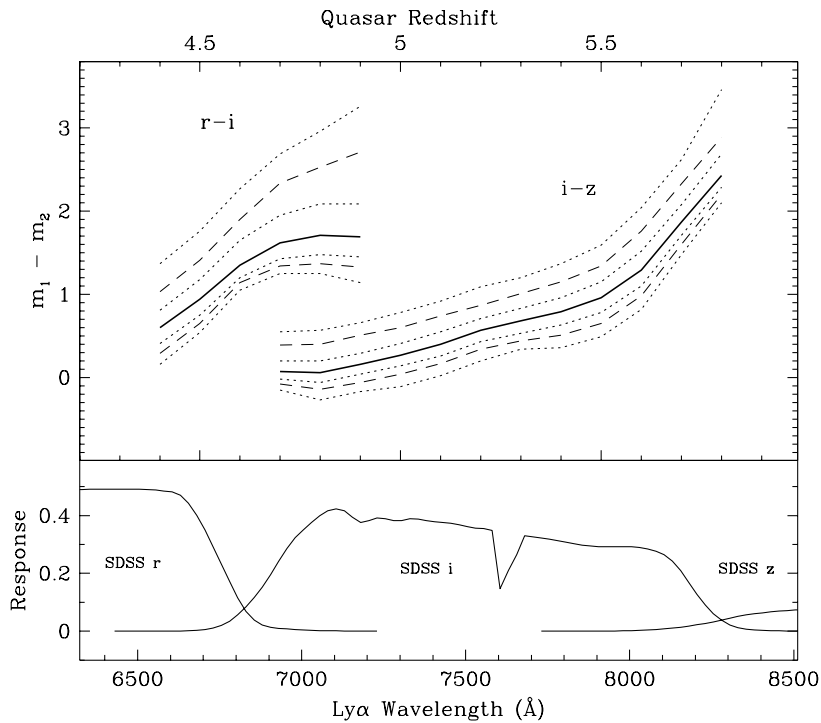}
\vspace*{-2.3in}
\caption{$r-i$ and $i-z$ median and 1\%, 5\%, 15\%, 85\%, 95\%, 99\% values of 741 cloned quasars at redshifts $4.4\le z \le 5.8$.  Also shown are SDSS $riz$ filter curves, with abscissa wavelength scale matching quasar redshift (Lyman-$\alpha$ break). Magnitudes are expressed in the SDSS {\it asinh} (AB) system.  The strong 7600\AA\, absorption in the SDSS $i$ filter is due to the atmospheric A band.}
  \end{center}
\end{figure*}

Each original SDSS spectrum was first smoothed by 5 pixels, then redshifted to the range of interest in 0.1 redshift bins.  During the redshifting procedure, the quasars were examined for the presence of flux at the extreme wavelengths of the filter set being tested.   For those quasars which were not complete down to the bluest filter or up to the reddest filter after redshifting, the data were extended synthetically by fitting and continuing the nearby quasar continuum level.  For example, 73\% of the quasars at $z\sim6$ required some extension of the data.  But on average only 200 pixels of data were added to the $\sim 3800$ pixels in each original SDSS spectrum.  This technique allowed a larger number of quasars to be used, with a larger redshift range.  Without such data extension, only quasars from much narrower ranges of redshifts would have been suitable for cloning, narrowing the diversity of spectral features.

The effect of Lyman-$\alpha$ absorption at high redshift was added by computing the difference in absorber optical depth between the redshift of the original quasar and the desired cloned quasar, adapted from the opacity behavior described by \citet{haardt}.  The appropriate average decrement was applied to each wavelength in the original data below the rest wavelength Lyman-$\alpha$.  We neglected the contribution from Lyman limit systems -- at all redshifts in these calculations, the rest wavelength $912$\AA\, fell outside the filters of relevance.  Finally, at redshifts $z>6.0$, complete Lyman-$\alpha$ damping was applied to the data -- {\it i.e.} Gunn-Peterson absorption \citep{gunn65}.

Finally, the synthetic magnitudes of each spectrum were computed in the relevant filters.   The SDSS $ugriz$ filter response functions are publicly available at http://www.sdss.org, while the UKIDSS response functions (not yet publicly available) were generated by modeling the imaging system.   Filter curves for the new UKIDSS {\it ZYJHK} filters were obtained from http://www.ukidss.org, and multiplied by typical HAWAII-2 HgCdTe detector efficiency curves and the near infrared atmospheric transmission function produced by the ATRAN program for average conditions (Lord 1992).  While computation of the absolute system efficiency requires additional multiplication by mirror reflections and secondary optical effects, these further steps were not applied because broadband colors are not affected by wavelength-independent factors. 

\begin{figure*}[t]
  \begin{center}
    \leavevmode
      \epsfxsize=12cm\epsfbox{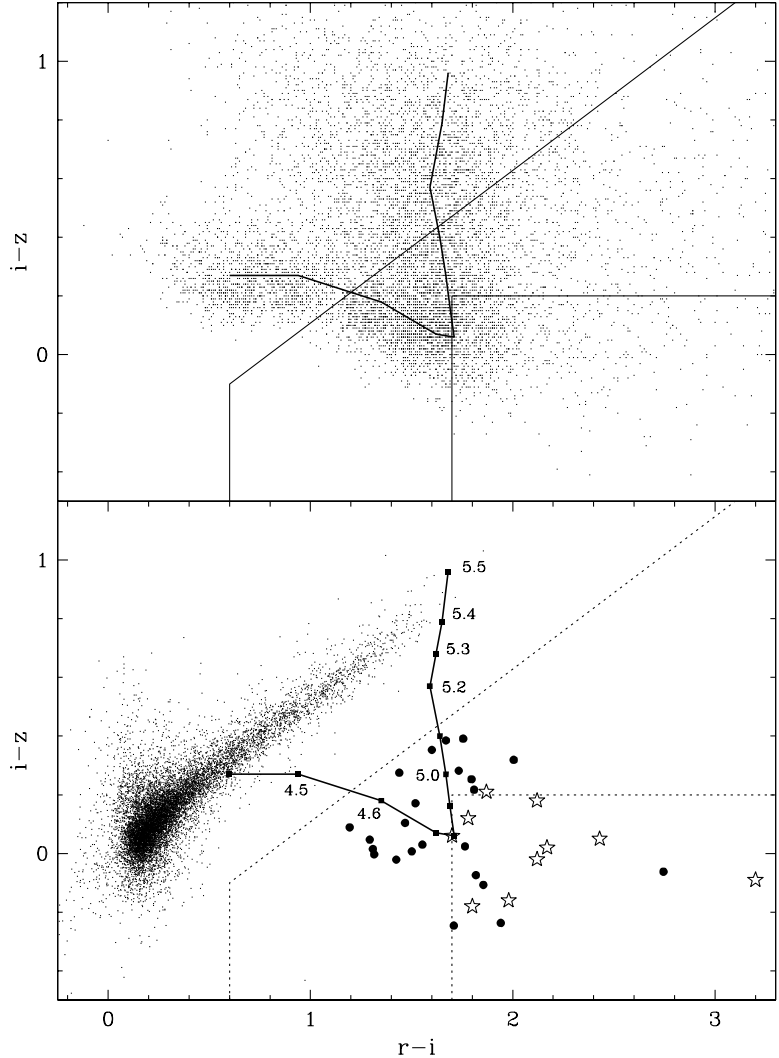}
      \vspace*{-0.3in}
       \caption{Top panel: 741 quasars cloned to 15 redshift bins at $4.4 \le z \le 5.8$ are plotted in ($r-i$, $i-z$) 2-color space.  Also shown is the SDSS hi-$z$ quasar selection function (slanted top) and selection function used in this work (smaller rectangular area).  Bottom panel: The quasar median track has been condensed (dark line with redshift labels), plotted with SDSS stellar locus (small points).  24 $z>4.7$ quasars found in SDSS DR1 survey (large dots) and ten new $z>4.7$ quasars discovered in this work (open stars) are shown.  Selection regions (dotted lines) are identical to top panel.  }
  \end{center}
\end{figure*}

The synthetic magnitudes were calculated according to the AB magnitude system described by \citet{oke}, and illustrated in \citet{blanton}:  
$$m_{AB} = -2.41 - 2.5\log _{10} \Biggl [ { \int_{0}^{\infty} d\lambda \lambda f(\lambda)R(\lambda)  \over  \int_{0}^{\infty} d\lambda \lambda^{-1} R(\lambda)  } \Biggr].$$  
Although optical AB magnitudes are used by the SDSS, infrared photometry is frequently expressed in Vega-based (Johnson) magnitudes -- in this work we employ both systems for ease of use by observers and note that an additive factor is needed to convert between the two.  For reference, the appropriate transformation between AB and Vega magnitudes is given by the equation: 
$m_{Vega} = m_{AB} + c, $ where $c$ is an additive correction dependent on wavelength and spectral energy distribution, we used $c=[-0.50,-0.64,-0.90,-1.38, -1.84]$ in the five UKIDSS filters $[Z,Y,J,H,K]$ based on a model quasar continuum of $f_\nu\propto\nu^{-0.5}$.  

In addition to the colors of the individual quasars, statistics of the samples versus redshift were compiled.   In all of the color versus redshift plots which follow, we show the median (50\%) 
quasar track accompanied by 1\%, 5\%, 15\%, 85\%, 95\%, and 99\% values.  

\section{Quasars at $z\sim 5$}

The $z\sim 5$ epoch has been a highly productive area for quasar discovery in recent years, with most of the advances originating from SDSS imaging data.  As shown in Figure 1 (bottom panel), quasars at these redshifts are identified in the SDSS $riz$ filters, and as seen in Figure 2 (bottom panel), are successfully targeted because of their separation from the stellar locus in 2-color space.  

However, not all quasars are observable due to the practical issue of stellar contamination -- and in fact it is this which imposes the upper redshift limit on the SDSS automated quasar spectroscopic program.  While the survey is in principle sensitive to quasars up to redshift $z=5.8$, stars overwhelm any color-selected observations beginning at $z=5.4$.  Only two quasars have been discovered in the broad range $5.4<z<5.8$ due to this problem (Stern et al. 2000, Romani et al. 2004).  Any successful quasar search based on colors must avoid such contaminating sources, which drive down spectroscopic followup efficiency -- therefore at certain redshifts such quasar programs will be necessarily incomplete.   

\subsection{Quasar Completeness at $z\sim 5$}

A primary goal of our quasar simulation work was to improve the completeness estimates of the SDSS quasar discovery rate.  There are only a few dozen known quasars at $z\sim5$, which places large errors on any statistics and completeness estimates.   With 741 cloned quasars, we now can estimate the SDSS selection completeness with higher precision.  

For the 741 quasars cloned to simulate the quasar track from $z=4.4$ to $z=5.8$, we have plotted the median $r-i$ and $i-z$ colors versus redshift, as well as the entire sample ($741\times 15$ objects) in $(r-i,i-z)$ to show the width of the distribution (Figure 2, top panel).   Also plotted are two quasar selection areas -- the first (slanted top) is that of the SDSS high redshift quasar program, defined by the boundaries $r-i>0.6$, $i-z>-1.0$, and $i-z<0.52*(r-i)-0.412$.  The smaller box is the selection used later in this work, $r-i>1.7$ and $i-z<0.2$, which is explained in the following section.  In Figure 2, lower panel, the $riz$ stellar locus and our median quasar track have been plotted for comparison.  

For each redshift from 4.4 to 5.5, we then calculated the fraction of the 741 cloned quasars falling in the SDSS selection, as well as in our smaller selection region.  The SDSS hi-$z$ quasar target selection is very successful, with a peak efficiency (quasar fraction captured)  of 98.2\% at redshift $z=4.7$.  Our smaller selection region, by comparison, has a peak of 41\% at $z=4.8$ -- but despite its lower absolute efficiency, its utility is illustrated in the accompanying observations we undertook.   These efficiency curves are shown in Figure 3, top panel.

\begin{figure}
  
    \leavevmode
      \epsfxsize=12cm\epsfbox{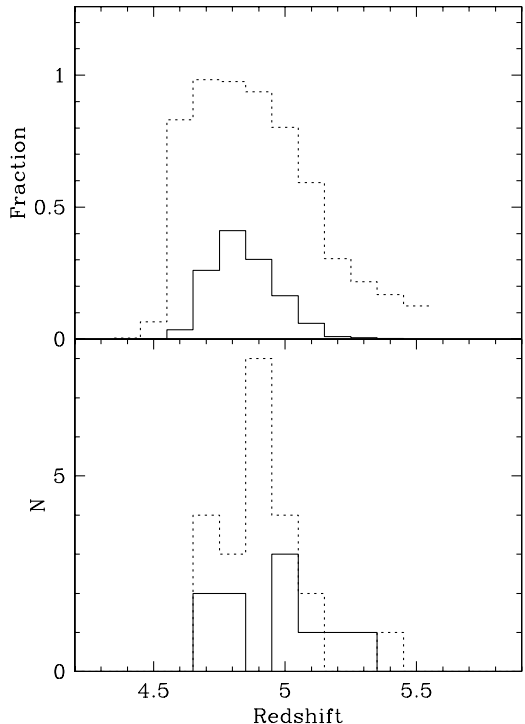}
      \vspace*{-2.2in}
       \caption{Top panel: $z\sim 5$ selection efficiency versus redshift, calculated as fraction of cloned quasars falling within two different selection criteria.  Dotted line is SDSS automated spectroscopic $z\sim 5$ selection, solid line is selection used by this work (as shown in Figure 2).  Lower panel: redshift distribution of quasars found.  Dotted line: 24 SDSS DR1 quasars with $z>4.7$, solid line: ten $z>4.7$ quasars found in this work.}
 
\end{figure}

\begin{table*}
\caption{Properties of ten $z>4.7$ quasars found in this work}
\begin{center}
\leavevmode
\begin{tabular*}{5in}{@{\extracolsep{\fill}}ccccccc}
\hline \hline 
object & redshift & {\it r} & {\it i} & {\it z} \\ 
\hline
J013326.84+010637.7 &  $5.30 \pm 0.01$    & $22.56 \pm 0.45$ & $20.69 \pm 0.05$ & $20.48 \pm 0.16$  \\
                    &       	          & $23.11 \pm 0.29$ & $20.72 \pm 0.06$ & $20.40 \pm 0.18$  \\
J074653.44+470517.6 &  $4.84 \pm 0.01$    & $22.32 \pm 0.17$ & $20.20 \pm 0.05$ & $20.22 \pm 0.17$  \\
J083317.66+272629.0 &  $5.02 \pm 0.01$    & $22.42 \pm 0.16$ & $20.25 \pm 0.04$ & $20.23 \pm 0.12$  \\
J084802.82+342715.3 &  $4.73 \pm 0.01$    & $22.65 \pm 0.16$ & $20.87 \pm 0.05$ & $20.75 \pm 0.15$  \\
                    &       	          & $22.94 \pm 0.36$ & $20.94 \pm 0.09$ & $20.93 \pm 0.30$  \\
J090059.51+274557.6 &  $4.96 \pm 0.01$    & $23.00 \pm 0.28$ & $20.57 \pm 0.05$ & $20.52 \pm 0.14$   \\
J134141.46+461110.3 &  $5.01 \pm 0.01$    & $22.19 \pm 0.13$ & $20.21 \pm 0.04$ & $20.37 \pm 0.15$  \\
J141026.22+385652.6 &  $4.75 \pm 0.01$    & $22.03 \pm 0.11$ & $20.23 \pm 0.04$ & $20.41 \pm 0.15$  \\
J152404.10+081639.3 &  $5.08 \pm 0.01$    & $22.81 \pm 0.20$ & $20.69 \pm 0.04$ & $20.51 \pm 0.11$   \\
J155422.97+303214.4 &  $4.84 \pm 0.01$    & $23.84 \pm 0.45$ & $20.64 \pm 0.05$ & $20.73 \pm 0.18$   \\
J211928.32+102906.6 &  $5.18 \pm 0.01$    & $22.32 \pm 0.12$ & $20.62 \pm 0.04$ & $20.56 \pm 0.15$  \\
                    &                     & $22.48 \pm 0.15$ & $20.54 \pm 0.04$ & $20.20 \pm 0.13$  \\
\end{tabular*}
\end{center}
\begin{center}
Double entries indicate quasar detected in more than one imaging run.
\end{center}
\end{table*}

\begin{figure*}
  \begin{center}
    \leavevmode
      \epsfxsize=13cm\epsfbox{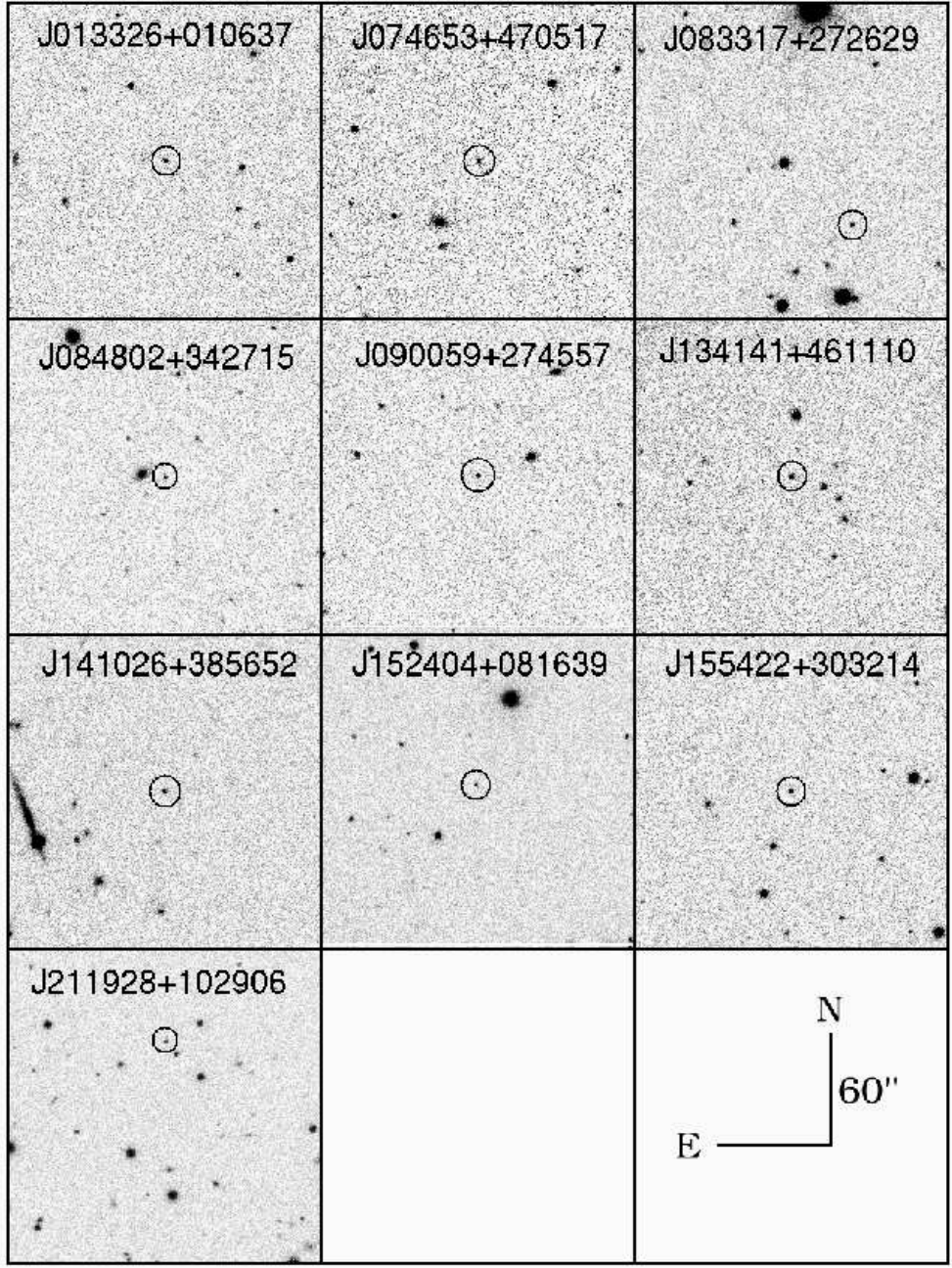}
      \vspace*{-0.0in}
       \caption{Finding charts for ten $z>4.7$ quasars found in this work, imaged in SDSS $i$ band.  North is up, East is left.  Boxes are 2 arcmin on a side.  }
  \end{center}
\end{figure*}

\begin{figure*}
  \begin{center}
    \leavevmode
      \epsfxsize=15cm\epsfbox{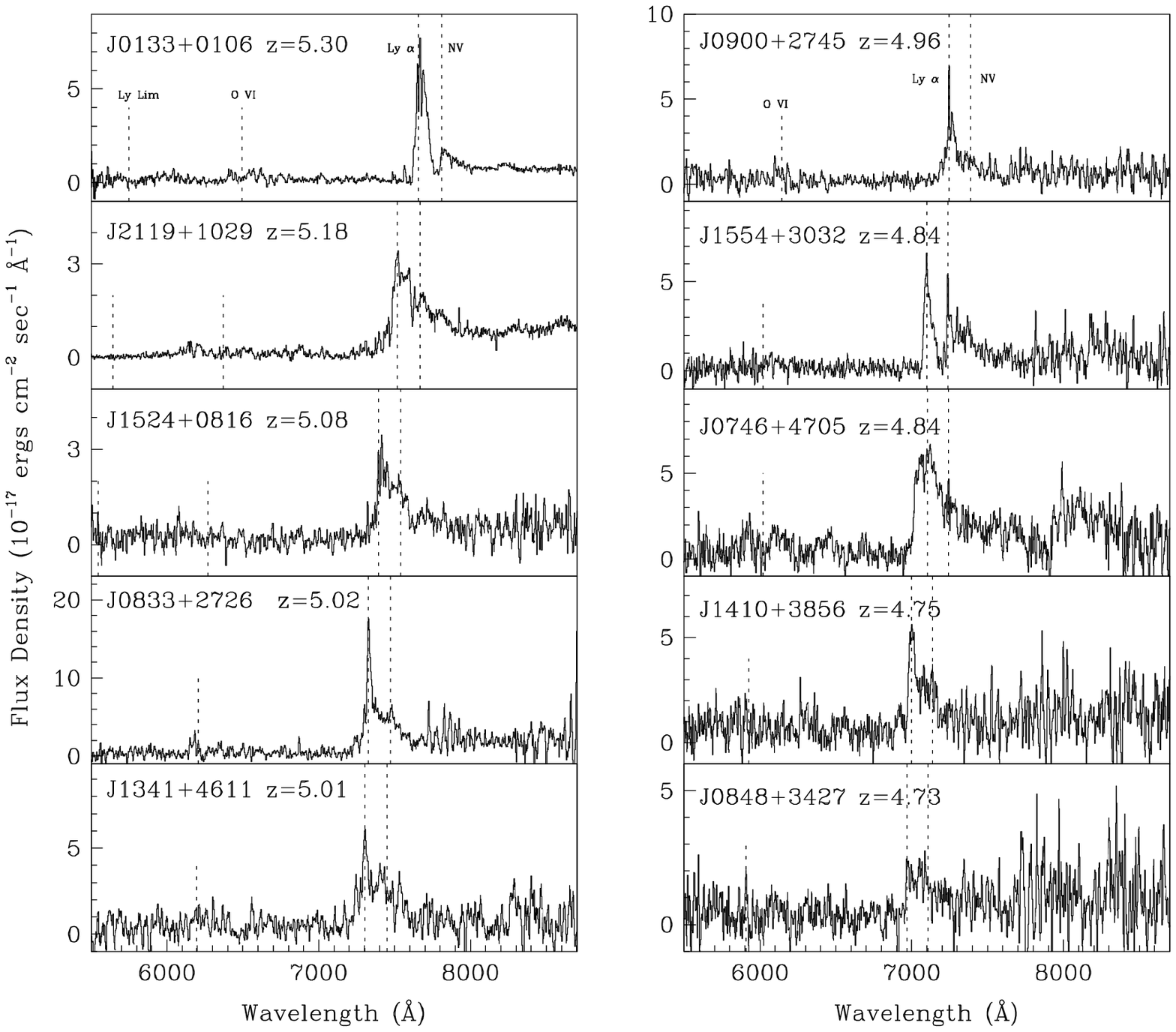}
      \vspace*{-2.9in}
       \caption{Spectra of the ten $z>4.7$ quasars found in this work. Peaks of hydrogen Lyman, NV, and OVI emission lines are marked.  }
  \end{center}
\end{figure*}

\subsection{Observing Quasars at $z\sim 5$: 10 Newly Discovered Objects}

We used these completeness results to design an observing program which successfully targeted faint $z\sim5$ quasars with high efficiency.  The observational challenge of finding $z\sim5$ quasars is relatively straightforward, though time intensive -- it involves selecting a suitable color sample of candidates, eliminating the bulk of contaminants, and carrying out spectroscopy.  With such a sample of faint quasars, we hoped to estimate the potential discoverable quasar pool in the SDSS imaging and make predictions about the future quasar discovery rates.   Other attributes of such a faint program are a significant increase in quasar candidate numbers and a larger resulting population of confirmed quasars.  In addition, a fainter quasar candidate selection can be expected to yield a population of slightly higher redshift objects than a brighter sample, due to the intrinsic luminosity properties of quasars. 

Using SDSS imaging data observed from September 2000 to March 2004, we performed the candidate selection, consisting of the aforementioned color cut and a magnitude cut on point source objects identified by the SDSS photometric pipeline, with precautions against false detections.  For the purpose of identifying quasars at $z\sim5$, we safely restricted our sample to point sources only -- of the quasars discovered automatically by the SDSS DR1 quasar survey, only 5.4\% of the total sample (913 of 17613) are classified as extended.  Above a redshift of $z=2.0$, the proportion falls more dramatically -- extended sources represent only 24 out of 3218 quasars.  By $z=4.0$ and greater, only 1 quasar out of 187 is classified as extended. 

We applied the requirement for $u$ and $g$ blank detections to the data, while also limiting the $i$-magnitude selection to $20.2<i<21.0$.  The {\it asinh} modified AB magnitude system of Lupton et al. (1999) produces reasonable magnitudes even at extremely low instrumental flux levels, which is critical for work near the survey flux limits.  $20.2<i<21.0$  represents the magnitude range unexamined by the SDSS spectroscopic target selection, down to the faintness limit of our spectroscopic capability (Richards et al. 2002).   The color cut, $r-i>1.7$ and $i-z<0.2$, was designed to produce a higher average redshift sample than could be expected from the SDSS main survey quasar selection, while decreasing the fraction of stars and other contaminants in the followup observations. 

A final measure ensuring high discovery efficiency during spectroscopic followup was the application of SDSS photometric flag screens to eliminate objects tainted by faulty photometry.  The simple application of color cuts alone often results in the retrieval of objects or artifacts dominated by cosmic rays, chip artifacts, moving sources, and other photometric errors.  Judicious application of flag requirements (such as excluding {\tt BRIGHT, EDGE, SATURATED, CHILD, CR}) eliminates most blatant problems while not excluding real sources.  Cosmic ray hits represented the bulk of remaining contaminants in our samples -- these were additionally removed through visual inspection.  

The candidates were observed using at the Double Imaging Spectrograph at the Apache Point Observatory (APO) 3.5m Telescope, the FOCAS instrument \citep{kashikawa} of the Subaru Telescope \citep{iye} at Mauna Kea, and the Low-Resolution Spectrograph \citep{hill} of the Hobby-Eberly Telescope.   Observational notes for individual objects are detailed here, and finding charts and spectra appear in Figures 4 and 5.  The typical exposure time was 1200 seconds with APO, or 300 seconds with Subaru, and spectral resolution was 2.5\AA\, over the range 5200\AA\, to 9500\AA\,.  In the case of quasar J084802.82+342715.3 ($z=4.73$), the intrinsic faintness of the object ($i=20.9$) combined with moderate to poor seeing and a minor tracking problem reduced flux.  Redshifts were determined by the SPECFIT program, with a fit of each smoothed quasar spectrum by a continuum, NV and Ly$\alpha$ emission, and absorption blueward of Ly$\alpha$.  

\begin{figure*}
  \begin{center}
    \leavevmode
      \epsfxsize=14cm\epsfbox{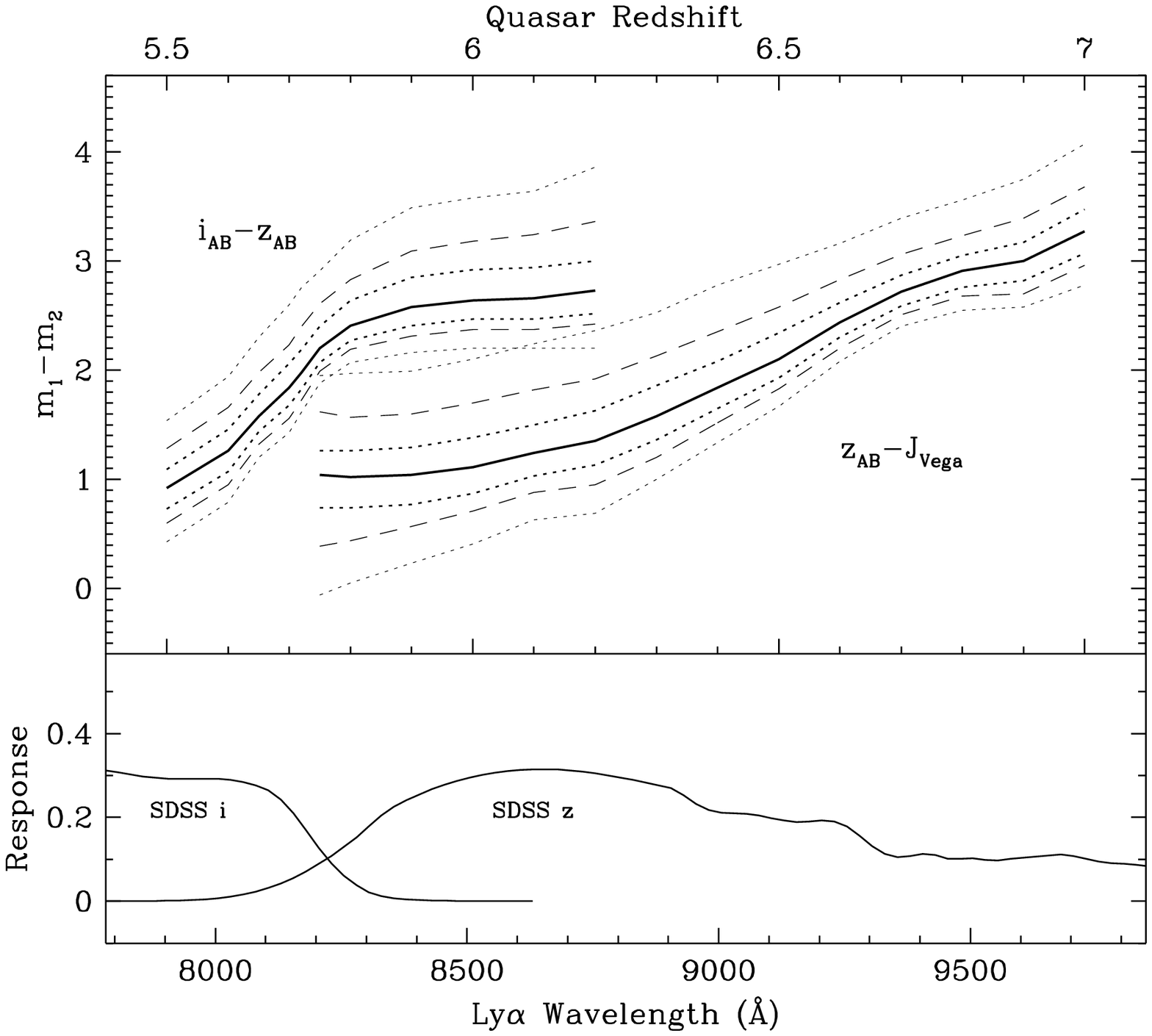}
      \vspace*{-2.8in}
\caption{$i-z$ and $z-J$ median and 1\%, 5\%, 15\%, 85\%, 95\%, 99\% values of 756 cloned quasars at redshifts $5.5\le z \le 7.0$.  Also shown are SDSS $iz$ filter curves, indicating location of quasar Lyman-$\alpha$ corresponding to redshifts. Magnitudes in this figure are mixed -- SDSS $i$ and $z$ are calculated in the SDSS AB system while UKIDSS $J$ is expressed in Vega system for ease of observers.  SDSS $z$ response has been multiplied by 4 to exaggerate shape. } 
  \end{center}
\end{figure*}

\begin{figure*}
  \begin{center}
    \leavevmode
      \epsfxsize=12cm\epsfbox{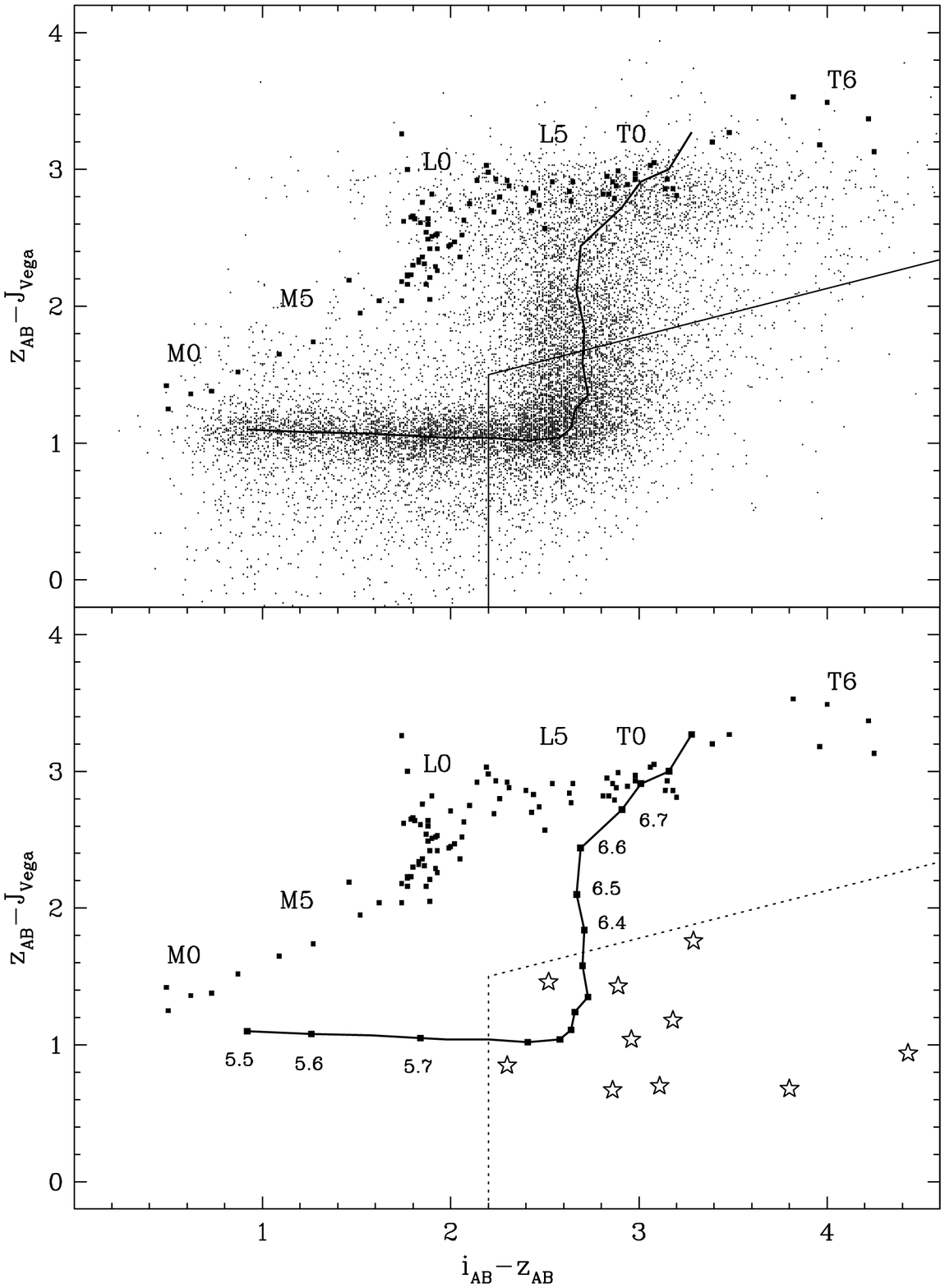}
      \vspace*{-0.3in}
\caption{Top panel: 756 quasars cloned to 16 redshift bins at $5.5\le z \le 7.0 $ are plotted in ($i-z$, $z-J$) 2-color space.  Also shown is the quasar selection function of Fan et al., $i-z>2.2$, $z-J < 1.5 + 0.35*(i-z-2.2)$.  The dwarf stellar locus (squares) is traced by known objects as reported in Hawley et al. (2002), Leggett et al. (2002), and Knapp et al. (2004).  Bottom panel: The quasar median track has been condensed (dark line with redshift labels), plotted with dwarf locus as in top panel.  10 $z>5.8$ quasars found to date by Fan et al. are shown (open stars).  Selection region (dotted line) is identical to top panel.  } 
  \end{center}
\end{figure*}

\subsection{Discussion of Sky Coverage and Statistics}

A comparison of these 10 quasars with quasars from the SDSS DR1 Quasar Catalog reveals that, as expected, a deeper spectroscopic observing program increases discovered quasar numbers by several fold.  We evaluate and quantify these gains here.  

The ten quasars described in this work were found in a searched sky coverage of approximately 656 square degrees.  The SDSS imaging scheme causes some sky area to be multiply imaged, and it is therefore necessary to calculate the total area searched as the unique footprint of the combined imaging runs, rather than their simple sum.  The simple overlapping areas searched in this work made such a calculation straightforward, but a more sophisticated calculation, such as Monte Carlo sphere filling, will be appropriate in future samples covering substantial and non-trivially overlapping areas.  Specifically, it must account for the 10\% overlap between strips and curved converging areas near the beginnings and ends of complete stripes at the survey poles (such as in Fan et al. 2001a).  In this analysis, such effects are small compared to the $\sqrt{10}$ Poisson noise of the sample size.  

As with the quasar cloning procedure, we use as a reference and comparison work the SDSS DR1 Quasar Catalog.  Of relevance to this analysis are quasars produced by the SDSS high-$z$ quasar target selection, which automatically selects targets expected to be of redshift $z>4.5$ from SDSS photometric data.  The DR1 contains 35 such quasars, from redshifts $z=4.56$ to $z=5.41$ with magnitudes from $i=18.33$ to $i=20.47$.   19 of these quasars have redshifts above $z=4.7$ and $i$-magnitudes within the SDSS spectroscopic survey limit of $i=20.2$.

We find that our sample of ten $z\sim5$ quasars is statistically consistent with the DR1 population of high redshift quasars.  In Figure 2, we overlay in ($r-i$, $i-z$) 2-color space the 24 DR1 quasars with redshift greater than 4.7 along with the ten quasars found in this work.  Also shown are our color selection criteria (rectangular box: $r-i>1.7$, $i-z<0.2$) and the SDSS high-z selection criteria (slanted top: $r-i>0.6$, $i-z>-1.0$, $i-z<0.52*(r-i)-0.412$).  Now examining only quasars which would be selected by both our and the SDSS quasar color criteria, we find seven DR1 objects for comparison which range from redshifts $z=4.69$ to $z=5.12$ and magnitudes $i=18.63$ to $i=20.45$.  Using the relative areas of each sample to compare the confirmed quasars in this specific color selection (1360 vs. 656 square degrees), we find that our sample produces $2.96\times$ the number of DR1 quasars per normalized area.  We attribute this gain to the fainter magnitude limit of our work, which reaches 0.7 $i$-magnitudes deeper than the SDSS main survey.   This result accords well with the work of \citet{fan01b}, for example, who described a number density increase of factor $\sim4$ per unit magnitude, for quasars at $z>4.0$.  

Treating the redshift range from $z=4.7$ to $z=5.2$ as one bin, we find a color selection completeness of 20\% in our work versus 76\% in the SDSS target selection by this measure.  Quasars selected by our target selection region represent 26\% of the sample selected by the SDSS quasar targeting algorithm.  This factor is not statistically very different from the 7 quasars out of 24 (29\%) found at redshifts $z>4.7$ in the DR1 Quasar Catalog.  

\subsection{Additional Object Discovery Rate}

A perfectly complete quasar luminosity function would specify quasar numbers at all redshifts, at all intrinsic brightnesses, with low statistical errors.  In reality, however, the faintness limits of observations and the natural rarity of quasars force a slightly less ideal specification of the luminosity function.  Extending the quasar luminosity function to high redshift and faint magnitudes is thus a primary goal of large area quasar surveys.  Recent work such as from 2QZ and the Sloan Digital Sky Survey (Croom et al. 2004, Schneider et al. 2003) have extended the knowledge of quasar density down to $M_{b_J}=-22$ and up to redshift $z=2.3$ and continue to press farther and fainter with deep spectroscopic followup observations.   

Although the $z\sim5$ quasar luminosity function will be poorly populated for some time, we can still estimate the numbers of quasars that can be found in future searches, using information from cloned quasar colors and the 10 quasars found in this work.   Given the SDSS ultimate goal of 10,000 square degrees of imaging, we expect a total of roughly 140 quasars eventually to be found with redshift $z>4.7$ and magnitude $i<20.2$, out of an ideal sample of 182 which in principle could be observed by the SDSS spectrographs if not for stellar contamination.  Further pursuit of faint quasars down to $i\sim20.9$ such as selected in this work would yield by the same calculation a total of 109 objects from our smaller color selection alone, or 414 in the larger SDSS hi-$z$ color selection region.  

One goal of the SDSS is to release deep imaging of the southern equatorial region, created from co-adds of areas multiply imaged over several years.  These regions have been imaged variously from $3\times$ to $18\times$ the normal survey exposure time, and an imaging depth of up to roughly $2.5 \log \sqrt{18} =1.56 $ magnitudes is expected.  From our limited sample here, we predict that such a depth increase could yield a quasar discovery rate up to $\sim 11\times$ that of the SDSS main survey.  We anticipate that followup spectroscopy within these deep areas will provide interesting tests of the quasar statistics and magnitude-number relations presented here, as well as shed new and much needed  light on the high redshift quasar luminosity function.

\begin{figure}
  
    \leavevmode
      \epsfxsize=12cm\epsfbox{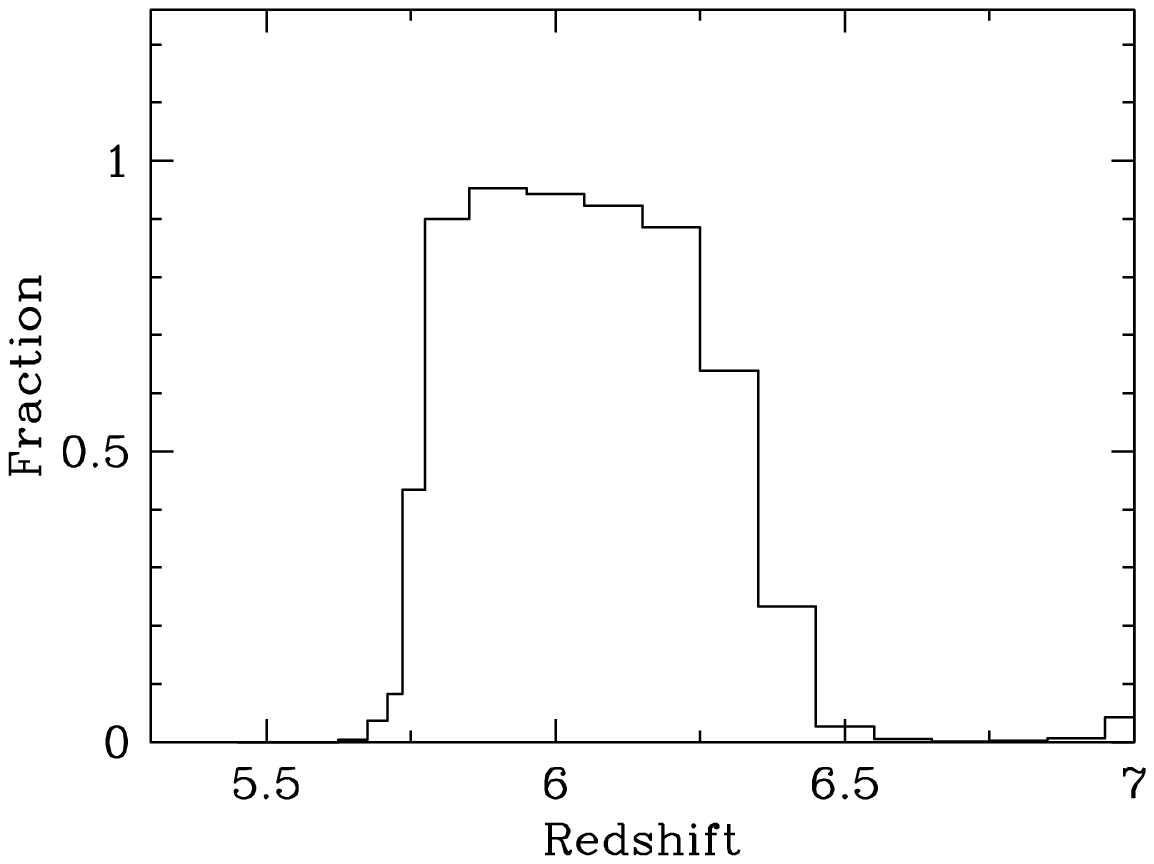}
      \vspace*{-4.in}
       \caption{$z\sim6$ selection efficiency versus redshift, calculated as fraction of cloned quasars falling within selection criteria of Fan et al. 2003 }
 
\end{figure}

\begin{figure}
  
    \leavevmode
      \epsfxsize=9cm\epsfbox{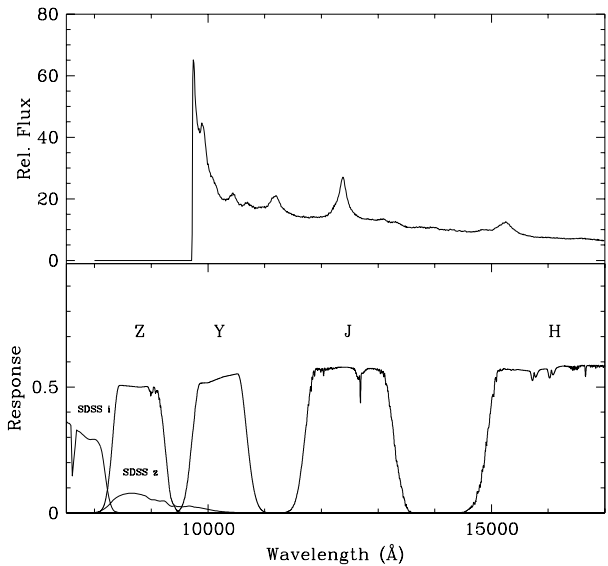}
      \vspace*{-1.6in}
\caption{Example of median cloned quasar at $z=7.0$, with strong emission features retained from original object spectra.  Flux is completely blacked out below Lyman-$\alpha$.  Shown in bottom panel are SDSS $i$ and $z$ filter responses, and new $ZY$ and $JHK$ filters of the UKIDSS.}
 
\end{figure}

\begin{figure*}
  \begin{center}
    \leavevmode
      \epsfxsize=15cm\epsfbox{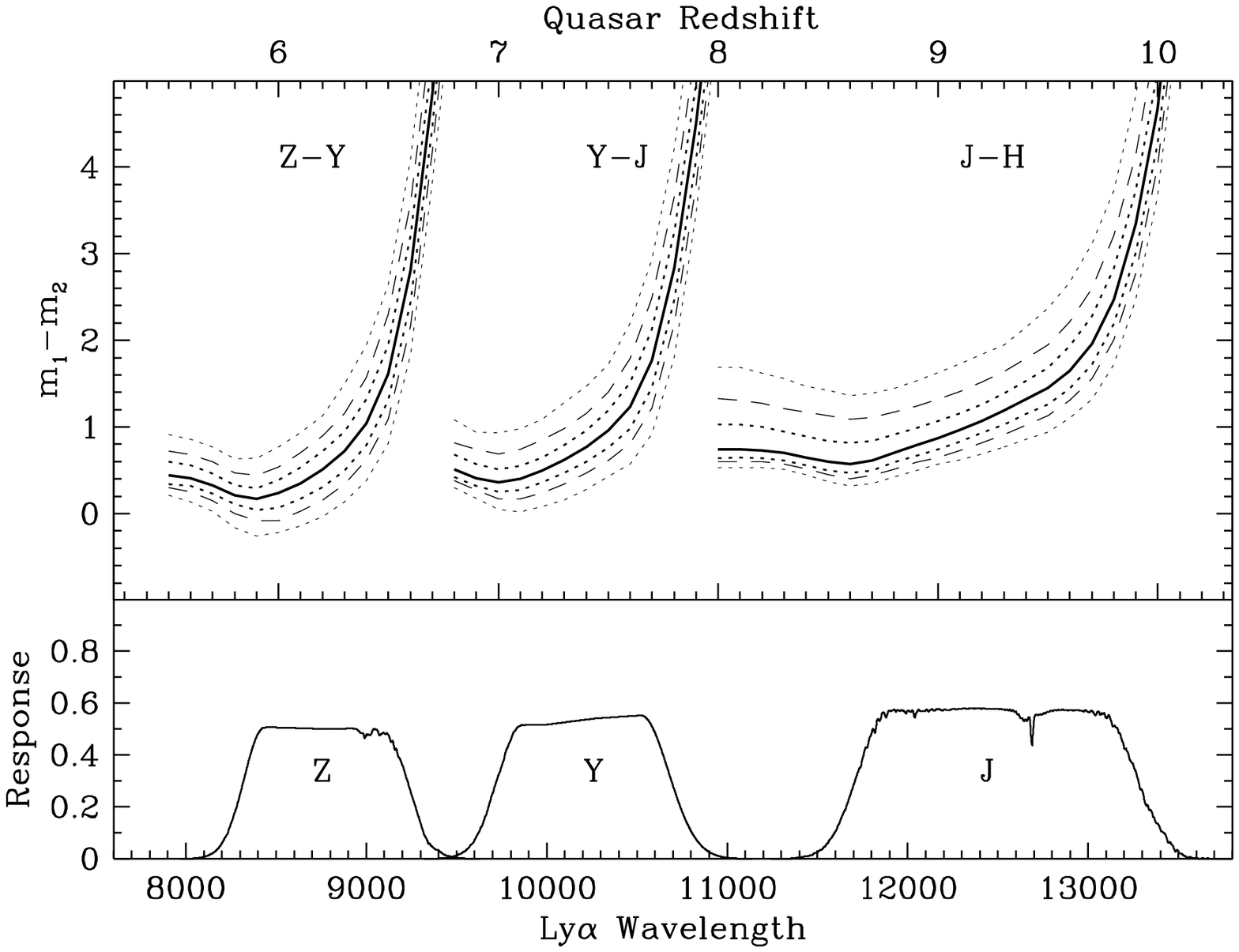}
      \vspace*{-3.7in}
\caption{UKIDSS $Z-Y$, $Y-J$, and $J-H$ median and 1\%, 5\%, 15\%, 85\%, 95\%, 99\% values of 1198 cloned quasars at redshifts $5.5 \le z \le 10.5$.  Also shown are {\it ZYJ} filter curves, indicating location of quasar Lyman-$\alpha$ corresponding to redshifts. Magnitudes in this figure expressed in Vega system for ease of observers. } 
  \end{center}
\end{figure*}

\section{Quasars at $z\sim 6$}

The SDSS imaging survey broke new ground by extending high quality imaging into the near infrared with the SDSS $z$ filter.  The relatively deep optical photometry, combined with followup infrared observations, allowed the discoveries of quasars for the first time at $z\sim 6$ (Fan et al. 2001a, 2003, 2004).  In a similar way to the quasars at $z=5$, quasars at $z=6$ are identified by their color separation from the contaminating stellar locus -- although in fact the ``contaminants'' of $z=6$ quasar searches are L- and T-dwarf stars, rare and interesting in their own right.  However, quasars at $z=6$ tax the observational stamina with the absolute rarity of objects as well as the challenge of flux in only one optical band: SDSS $z$.  Quasar candidates at $z>5.8$ are $i$-dropouts, and as high-risk single-band detections must be targeted and followed ``manually'' -- recent discoveries of quasars at $z\sim6$ have all been discovered through dedicated traditional spectroscopic followup rather than by an automated survey mode.  

To understand the completeness of the $z\sim6$ quasars discovered to date, we simulated the colors of 756 cloned quasars in the SDSS $iz$ and UKIDSS $J$ filters, according to the procedure introduced in the preceding sections.  Figure 6 shows the evolution of the $i-z$ and $z-J$ colors.  The sample begins at redshift $z=5.5$, where the quasar Ly$\alpha$ (now at $\sim 7900$\AA) has already been redshifted through much of the the SDSS $i$ filter.  As the quasars continue to redshift through the remaining portion of the $i$ band, measured $i$ flux drops rapidly while the Ly$\alpha$ peak quickly brings flux into the $z$ band. Thus within 0.3 redshift units, by $8300$\AA ($z=5.8$), the $i-z$ color has experienced the bulk of its increase -- from 0.92 to 2.41.   After this point the $i-z$ value saturates, with some small fluctuations as remaining flux blueward of Ly$\alpha$ moves through the $i$ band.   By contrast, due to the extreme width of the SDSS $z$ filter, the $z-J$ increase is gradual compared to the rapid rise in $i-z$, though ultimately similar in magnitude.   We follow the quasar $z-J$ value from $z=5.7$ until $z=7.0$, where the Ly$\alpha$ has reached $9700$\AA, nearly the red edge of the $z$ filter.

The $756\times 16=12096$ ($i-z,z-J$) points are shown in Figure 7 (top panel) along with the successful high-redshift quasar color selection region of Fan et al. (2001a, 2003, 2004) which chooses colors $i-z > 2.2$, and $z-J < 1.5+0.35*(i-z-2.2)$.  Also shown are dwarf stars from \citet{hawley}, \citet{leggett}, \citet{knapp}, tracing the now very red stellar locus.  In the bottom panel of Figure 7, the condensed median track of the 756 cloned quasars is plotted with labels showing redshift evolution, again plotted with the selection region of \citet{fan03}.  Overlaid are the 10 quasars at $z>5.8$ discovered by them over the past 3 years using this selection (starred symbols; one quasar without $J$ photometry is omitted). 

The quasar color selection used by Fan et al. is necessarily optimized for a particular redshift range.  Using the same procedure as detailed for the $z\sim5$ quasars, we computed the redshift of peak selection effectiveness, again defined by the fraction of cloned quasars selected at each redshift bin.  This is shown in Figure 8.   The peak of the Fan et al. selection occurs at $5.8<z<6.2$, with 90-95\% of objects selected, not accounting for contamination by stars.  Neither does this account for flux limitations of the survey, which tends to decrease the effectiveness of selection with higher redshifts. Efficiency decreases from the peak at $z=5.9$ to practically zero by redshifts 5.65 and 6.5.   

The $izJ$ quasar evolution track found by our simulations is very similar to that of \citet{fan03}, though some small deviations occur from $z=5.6$ to $z=5.9$ and at $z>6.3$.  The model quasar track of Fan et al. departs from our track gradually beginning at $z=5.5$, separating by a maximum of $z-J=0.3$ at $z=5.9$.  This difference falls within the first lower $z-J$ error surface at $z=5.9$, and we attribute it to strong emission line properties of our sample.   In Figure 9, we show a median $z=7.0$ quasar found in our cloned population.

\section{Quasars at $z>6.5$}

As quasars approach redshift $z=6.5$, the Sloan Digital Sky Survey reaches the limits of its observational capabilities.  Both the fading apparent magnitudes of quasars at this distance and the fading quantum efficiency of CCD imaging at 1$\mu {\rm m}$ conspire to hinder further discoveries of quasars at higher redshifts.  Thus the discovery of quasars at $z>6.5$ will mainly depend on ongoing deeper near-IR surveys.  

\subsection{{\it ZYJHK} and Future High Redshift Quasar Selection}

The UKIDSS {\it ZYJHK} filter system was developed in part to allow the next generation of quasars at $z>6.5$ to be discovered.  Most significantly, the novel $Z$ and $Y$ filters fill the large gap existing between the SDSS $z$ and near-IR $J$ bands with sharp-cutoff bandpasses.  This stretch of wavelength is of great importance for the discovery of $z>6.5$ quasars because the Lyman-$\alpha$ break must be observed straddling two adjacent bands in order to select targets with high certainty.  Quasars at $z=7.0$, for example, fall squarely in the gap between SDSS $z$ and near-IR $J$, but are rescued by the new $Y$ filter.  Combined with the high quantum efficiency of HAWAII HgCdTe detectors from 0.78 to $2.6{\rm \mu m}$, this system is typical of the future capability of infrared surveys.  

To date, one model of quasar colors in {\it ZYJH} has been advanced, namely in \citet{warren} -- so we now use the spectral cloning technique to bring real quasar spectral data to the problem.  1198 quasars from redshifts $2.3\le z\le 3.32$ were used for the cloning, and these were redshifted to the range $5.5\le z \le 10.5$.  As with the $riz$ and $izJ$ samples around $z\sim5$ and $z\sim6$, these quasars were selected in order to ensure the presence of original spectral information in the final cloned quasars throughout the bands {\it ZYJH} -- the filters of relevance to simulations up to $z=10$.  

The resulting color curves are shown in Figure 10.  For comparison, the location of the {\it ZYJ} filters are also shown, with both redshift and wavelength labels corresponding to the location of Lyman-$\alpha$.   

\begin{figure}
  
    \leavevmode
      \epsfxsize=16cm\epsfbox{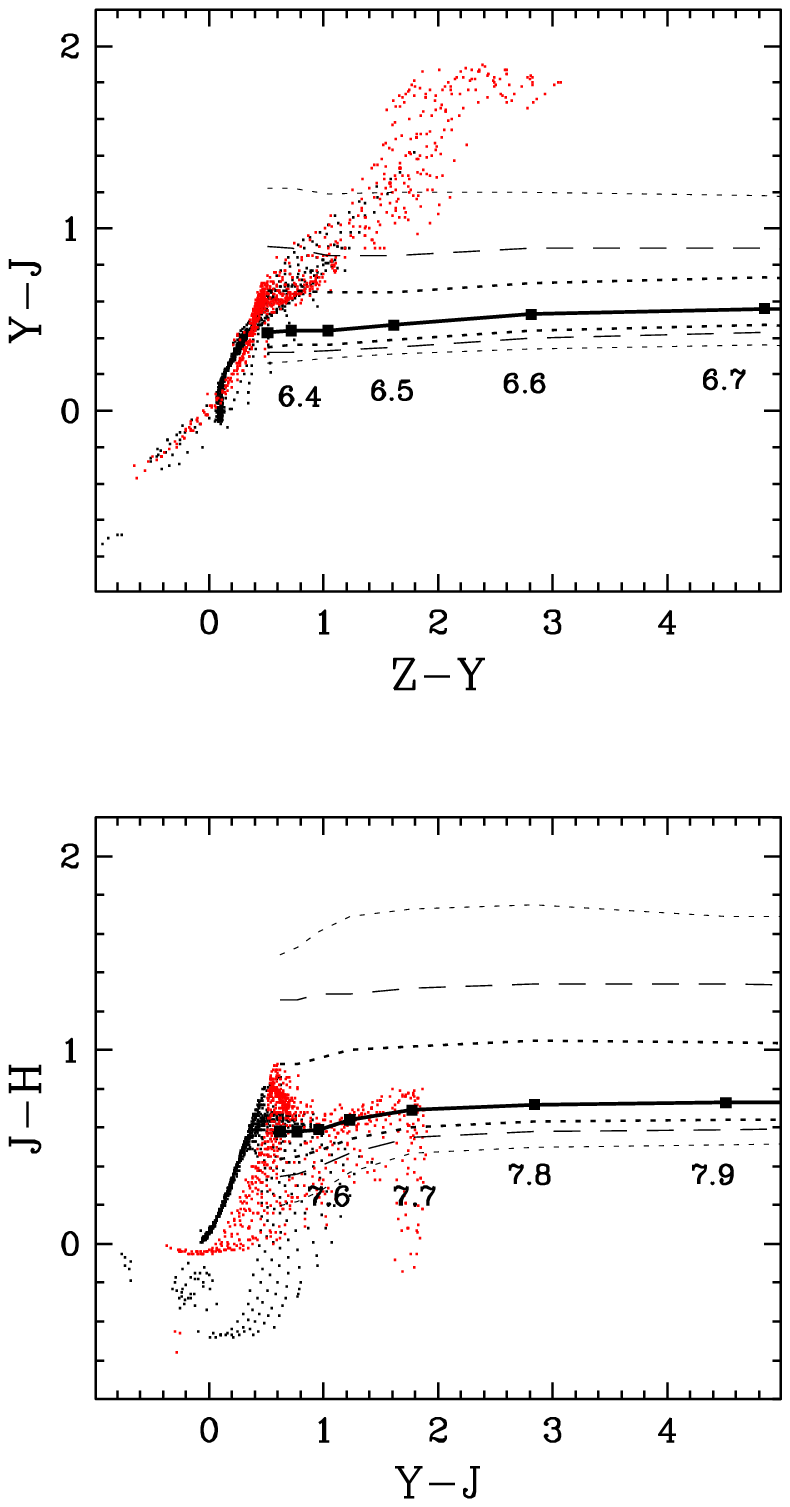}
      \vspace*{-3.2in}
       \caption{ 2-color plots of UKIDSS ($Z-Y$, $Y-J$) and ($Y-J$, $J-H$) median quasar tracks with 1\%, 5\%, 15\%, 85\%, 95\%, 99\% values.  Evolution of quasar track shown with labels indicating redshifts.  Dwarf stellar locus and normal stellar locus are plotted with small points, using model stellar spectra of Hauschildt et al. (1999) and Allard et al. (1995).  See the electronic edition of the Journal for color version. }
 
\end{figure}

\subsection{Quasar Colors at $z>6.5$}

The behavior of the three {\it ZYJH} color curves are very similar -- quasar color rises gradually as the Lyman-$\alpha$ break passes through the band, and increases rapidly as it leaves the tail of the filter.  This rapid rise is thus the most sensitive regime of the color-redshift measurement.  However, the three color curves presented in Figure 10 are an interesting departure from the curves of lower redshift quasars, in which small, but significant flux could usually be relied on to be present below Lyman-$\alpha$.  This minimal flux served to ensure that even after the passage of the Lyman-$\alpha$ break through an entire band, model spectra would have some flux remaining in the blueward band, and thus reasonably stable colors with increasing redshift -- {\it i.e.} on the order of $m_1-m_2 < 5$.

With quasars above $z=6.5$ however, we encounter for the first time (in the sense of survey planning) an expected {\it complete} damping of the residual flux, which leads to the possibility -- at least in modeled systems and synthetic spectra -- of arbitrarily large colors.  While we plot color values up to $m_1-m_2= 5$, in fact our simulations continue beyond this point to find values as high as 13 or 14 -- which are clearly not meaningful for observational purposes.  Because a value of $m_1-m_2>5$ is both consistent with Lyman-$\alpha$ only slightly redshifted beyond the blueward band, {\it and} Lyman-$\alpha$ having been well redshifted out of the blueward band, it is more reasonable to rely on the next color for redshift estimation and treat the first color as a lower redshift limit.  In 2-color diagrams of this redshift range (Figure 11) therefore, we remind that the redshift track should be regarded as degenerate above $x$-values of 5.

Finally, a major barrier to the potential success of {\it ZYJH} quasar color selection is the declining spatial density of objects with increasing redshift.  In this respect the difficulty of finding $z>6.5$ quasars is of course multiply compounded compared to $z\sim6$ -- the quasars are expected to be rarer, fainter, and of higher color decrement -- factors which all pose observational challenges.  If we adopt the decreasing quasar spatial density evolution of Fan et al. 2004 and predict $M_{1450}<-26.7$ quasars at even higher redshifts, we find that in the range $6.5\le z \le 7.5$, a total of only 90 such quasars are to be expected {\it on the whole sky}.  Quasars of this brightness, if they exist at redshifts $z\sim7$, will have moved into the $Y$ band and will be fainter by a half magnitude compared to the known objects at $z\sim6$.   The detection of $z\sim7$ quasars will be a challenging endeavor indeed.  

What choices will be faced in optimizing the likelihood of observing such rare, distant objects?  Clearly, the tradeoff between limiting magnitude and areal coverage is a significant concern for any large extragalactic survey.  Ordinarily, if a quasar or galaxy survey seeks to maximize the number of objects observed, the decision whether to pursue deeper observations at the cost of survey area is dependent on the targeted objects' luminosity function.  Specifically, for example, surveying new area with incremental observing time is generally expected to yield new objects in proportion to the additional time spent, i.e. $2\times$ area = $2\times$ objects.  And in order for this additional time to be spent as effectively on deeper observations instead (where $S/N\propto \sqrt{t}$), the objects of interest must obey a number density increase of at least $2^{({1\over 2.5\log\sqrt2})}=6.3\times$ per magnitude.  However, in surveys for the highest redshift objects, which naturally seek the few most luminous objects first, depth is a necessary precursor to wide areas.  At the bright tail of the luminosity function where objects brighter than a certain absolute magnitude are effectively non-existent, doubling the area will yield no objects if the survey is not deep enough.  In such cases then, appropriately deep, moderate- to smaller-area observations seem to most effectively use finite telescope resources.  Surveys such as UKIDSS, which will probe to limits of $Y=20.5$, seem to offer hope for establishing the luminosity properties of these desired rare objects, in preparation for casting a broader area search.  

\vspace*{0.3in}
\section{Conclusion}

Large-area sky surveys have often been motivated by the potential discovery of large numbers of objects which are just out of reach with present capabilities.  The rapid advance in quasar information achieved in the past several years through such surveys has provided the foundation for extending the redshift frontier even farther, with digital libraries of quasars that can be used to predict new, even more distant populations.  With a large number of quasars at low redshift, we have cloned and simulated the diverse range of colors of potentially observable objects at $z>6.5$, as well as shown how quasar cloning can improve the completeness estimates of sparse quasar samples at intermediate redshifts, such as $z\sim5$ and $z\sim6$.  

The ten quasars discovered at redshifts $4.73<z<5.30$ in this work illustrate how cloned quasar color simulations can aid in the search for faint objects, and help build more robust statistics of sparse populations.  We reached 0.7 magnitudes deeper than the SDSS spectroscopic survey with this targeted search, and project a $3\times$ quasar discovery rate for this depth increase -- deeper co-adds of SDSS imaging (planned for release) therefore seem to offer excellent prospects for additional discoveries of faint, distant quasars. 

As the focus of cosmological attention is increasingly shifted to longer wavelengths, observers will seek deeper and redder observations to probe the properties of early objects.  Aside from the spectacular depth gains that can be achieved by a space-based infrared survey mission such as {\it PRIME} \citep{zheng02}, UKIDSS and similar ground-based infrared surveys provide the best hope of glimpsing the $z\sim7$ epoch of quasar activity.  The diversity of objects simulated in this work we hope will provide a useful guide for the understanding and discovery of the next generation of quasars.  

\section{Acknowledgments}

The authors thank P. Hewett and S. Warren for providing useful information regarding the UKIDSS filter system.  This research was partially supported by NSF grant 03-07582 (DPS).  KG acknowledges support from the David and Lucille Packard Foundation.  Observations were carried out at the Apache Point Observatory 3.5-meter telescope, which is owned and operated by the Astrophysical Research Consortium, and at Subaru Telescope, which is operated by the National Astronomical Observatory of Japan, and at the Hobby-Eberly Telescope, a joint project of the University of Texas at Austin, the Pennsylvania State University,  Stanford University, Ludwig-Maximillians-Universit\"at M\"unchen, and Georg-August-Universit\"at G\"ottingen.  The HET is named in honor of its principal benefactors, William P. Hobby and Robert E. Eberly.   We thank the telescope and instrument staffs at the Subaru Telescope and the Apache Point Observatory for their expert assistance and knowledge during observations.  

  Funding for the creation and distribution of the SDSS Archive has
been provided by the Alfred P. Sloan Foundation, the Participating
Institutions, the National Aeronautics and Space Administration, the
National Science Foundation, the U.S. Department of Energy, the
Japanese Monbukagakusho, and the Max Planck Society.  The SDSS Web
site is http://www.sdss.org/.

The SDSS is managed by the Astrophysical Research Consortium (ARC) for
the Participating Institutions.  The Participating Institutions are
The University of Chicago, Fermilab, the Institute for Advanced Study,
the Japan Participation Group, The Johns Hopkins University, the
Korean Scientist Group, Los Alamos National Laboratory, the
Max-Planck-Institute for Astronomy (MPIA), the Max-Planck-Institute
for Astrophysics (MPA), New Mexico State University, University of
Pittsburgh, University of Portsmouth, Princeton University, the United
States Naval Observatory, and the University of Washington.

\end{document}